\documentclass[9pt,a4paper,twocolumn]{narms2}

\usepackage{subfigure} 
\usepackage{psfig}
\usepackage{epsfig} 
\usepackage{timesmt}   
\usepackage{chikako}   

\makeatletter

\renewcommand{\section}{\@startsection%
{section}%
{1}%
{0mm}%
{- \baselineskip}%
{0.15\baselineskip}%
{\normalfont\normalsize}}%

\renewcommand{\subsection}{\@startsection
{subsection}%
{2}%
{0mm}%
{-\baselineskip}%
{0.15\baselineskip}%
{\normalfont\normalsize}}%
\makeatother

\begin{document}

\title{Kinetic Monte-Carlo simulations of sintering}
\author{\large {F. Westerhoff \& R. Zinetullin \& D.E. Wolf}\\
{\em Fachbereich Physik, Universit\"at Duisburg-Essen, Germany}\\
}
\date{}

\abstract{
We simulate the sintering of particle aggregates due to
surface diffusion. As a method we use Kinetic Monte-Carlo simulations
in which elasticity can explicitly  be taken into account. Therefore
it is possible to investigate the shape relaxation of aggregates also
under the influence of an external pressure. 
Without elasticity we investigate the relaxation time and surface
evolution of sintering aggregates and compare the simulations with the
classical Koch-Friedlander theory. Deviations from the theoretical
predictions will be discussed.}


\maketitle
\frenchspacing   


\section{INTRODUCTION}
One interesting aspect of powder processing is thermal
sintering: Due to atomic diffusion solid bridges between particles
form and grow. This lowers the surface free energy. If one waits long
enough, complete coalescence of the particle aggregate will
occur. The typical time scales depend on the particle sizes and shapes, the
material type and the prevailing temperature. Here we 
address open questions concerning the sintering dynamics of large 
aggregates.  

It has long been known, that the time for the sintering of two identical
particles is proportional 
to the 4th power of the radius of the final particle. This is at least
true for large particles above the
roughening temperature \shortcite{NicholsMullins}. For lower
temperatures the equilibration time increases
exponentially \shortcite{CombeJensen}. 

In earlier investigations elasticity has not explicitly been taken into
account, i.e. the atoms were restricted to discrete lattice positions.
However, an elastic deformation changes the diffusion constant and hence
influences the sintering process. For example, the lattice constant of
nano-particles may differ from its bulk value due to surface tension,
or particles may be compressed differently in a powder under external
load, depending on their position in the contact force network.
Therefore we developed a program where the atoms are allowed to be
displaced from their lattice sites in order to minimize the total
elastic energy of the system.

\section{MODEL}
We use a three dimensional Kinetic Monte-Carlo (KMC) method to
simulate the sintering of aggregates of nano-particles on a fcc
lattice. This means that grain boundaries are neglected in this paper.
In our simulation the atomic displacements are calculated by finding
the nearest local energy minimum by means of a conjugate gradient
method \shortcite{NumericalRecipes}. In order to save computing time,
we update only the neighborhood of diffusing atoms, and relax the
whole system elastically only in the beginning and, if necessary,
again after rather large time intervals.  The activation energies for
the hopping rates then depend on the {\em relaxed} real positions of
the atoms.

The atoms interact via a Lennard-Jones
potential. We calculate the binding
energies $E_{\rm b,i}$
as the sum of the pair interactions up to the fourth nearest
neighbor on the underlying fcc-lattice . 

\begin{figure}[b]
  \centering                           
  \epsfig{file=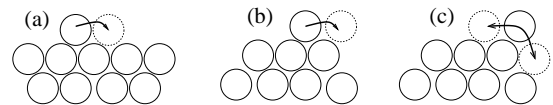,width=0.45\textwidth}      
  \caption{2d illustration for an atom that hops to a stable (a),
  respectively to an unstable intermediate position (b) from where it
  immediately continues to one of the neighboring stable sites (c).}
  \label{fig:hops}                        
\end{figure}

In a diffusion step an atom hops from its initial position (nominal
lattice site $i$) to an unoccupied one, which is next to $i$ on the
underlying fcc-lattice.  The activation energy for such a move is the
difference between the energy at the saddle point and the energy at
the initial position of the atom. For the calculation
of the saddle point energy $E_{\rm sp}$ one has to distinguish two
cases: The final site may be stable (at least three occupied neighbors) or
unstable. In the latter case one can actually not find a local energy
minimum at this site (see Fig.\ref{fig:hops}). Then the atom continues
to move from this intermediate site to a randomly chosen stable final
position next to it (with some probability it may actually jump back
to its starting position $i$).

\begin{figure}
  \centering                           
  \epsfig{file=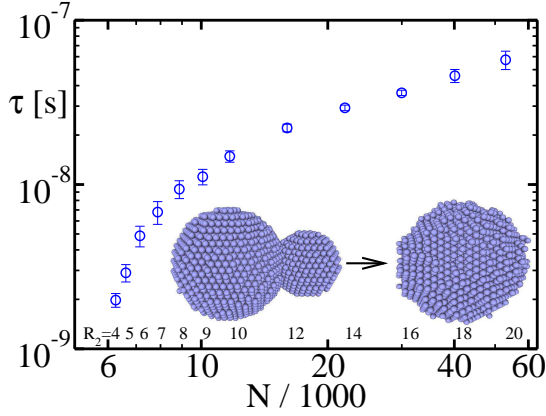,width=0.45\textwidth}      
  \caption{Equilibration time for two particles of different size
  plotted versus the number of atoms ($T=800$K). 
  The radius of particle
  one is 10 atoms, the radius for particle 2 is varied from 4 to 20
  as denoted in the bottom of the graph. ) }           
  \label{fig:teq_vs_N_twodiff}                        
\end{figure}                           

Accordingly the saddle point energy for the first case
is taken as
\begin{equation}
 E_{\rm sp} = E_{\rm sp,0} +\alpha\frac{1}{2}(E'_{\rm
    b,i}+E'_{\rm b,f}).     \label{eq:Ealj_stable}
\end{equation}
The first term is a constant parameter. The second one describes the
strain dependence of the saddle point energy. $E'_{\rm b,i}$ and
$E'_{\rm b,f}$  are the strain dependent contributions to the binding
energy at the initial (`i') and final (`f') atom
positions. The average is taken to guarantee symmetry and $\alpha$ is
an empirical parameter. This ansatz is justified by the finding that
the binding as well as the saddle point energy depend approximately
linearly on the strain.\shortcite{Schroeder}

If the hop ends in an unstable, intermediate position, this is
approximately regarded as the saddle point. We calculate the energy of the
atom at the intermediate site giving it the average displacement of 
the stable atoms on the neighboring sites. This energy is taken as
saddle point energy in the second case.

In the KMC simulation every process is selected according to its
corresponding rate which is calculated from the activation energy by:
\begin{equation}
q = \nu \, \exp{\left( -\beta E_a \right)}.
\end{equation} 
$\nu$ is a fixed attempt frequency, $\beta=1/k_{\rm B} T$ with the
Boltzmann factor $k_{\rm B}$ and the absolute temperature $T$.

Without elasticity $E_{b,i}$ corresponds to the bond counting
model\shortcite{Newman}  
and the strain dependent terms in (\ref{eq:Ealj_stable}) vanish.


\section{RESULTS}

\subsection{Relaxation of two clusters}

Crystalline particles above their roughening temperature are not
faceted but round. If they have equal size, it is well 
known that the equilibration time, in which the two particles coalesce
into a single one, is:
\begin{equation}
\tau \propto N^{4/3}\ ,
\label{eq:Nichols_Mullins}
\end{equation}
where $N \propto r^3$ is the total number of atoms.

\begin{figure}
  \centering                           
  \epsfig{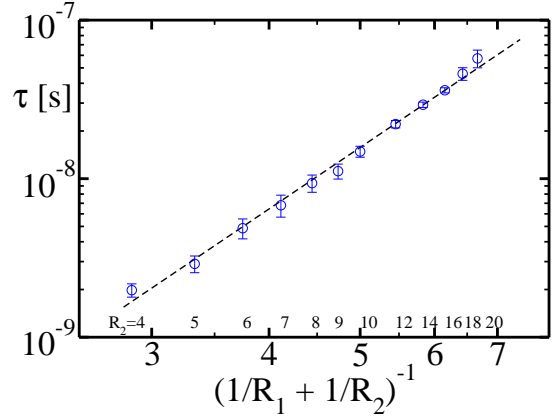}      
  \caption{The same data as in fig.~\ref{fig:teq_vs_N_twodiff} plotted
  versus the reduced radius.  }          
  \label{fig:teq_vs_Rred_twodiff}                        
\end{figure}

But what happens if the two diameters differ? In
Fig.~\ref{fig:teq_vs_N_twodiff} the equilibration time is plotted versus
the number of atoms. If (\ref{eq:Nichols_Mullins}) were valid, the
double logarithmic plot should give a straight line with slope $4/3$.   
Instead we find that $\tau \propto r^4$ with the reduced radius of the
two particles,
\begin{equation}
r = \left(\frac{1}{R_1} + \frac{1}{R_2}\right)^{-1}
\label{eq:Sinterzeit}
\end{equation}
(see Fig.~\ref{fig:teq_vs_Rred_twodiff}).

\subsection{Surface evolution during sintering}

The surface evolution during the sintering process is usually described
by the Koch-Friedlander theory \shortcite{Koch-Friedlander}:
\begin{equation}
\frac{dA}{dt} = - \frac{A - A_{\rm eq}}{\tau}.
\label{eq:Koch-Friedlander}
\end{equation}
Here $A(t)$ is the surface of the agglomerate, $A_{\rm eq}$ the
surface in equilibrium, $\tau$ a relaxation time. However, this
equation is interpreted and applied in several different ways in the
literature. Sometimes $\tau$ and $A_{\rm eq}$ are viewed as constants,
meaning the global relaxation time and the equilibrium surface area the
whole aggregate will have after coalescence. In this case
Eq.~(\ref{eq:Koch-Friedlander}) describes an exponential decay.

In many cases $\tau$ and
$A_{\rm eq}$ are regarded as time dependent, local quantities. 
One assumes that the aggregate coarsens homogeneously by pairwise
coalescence of particles. As explained above, the time constant $\tau$
is then determined by the current radius of the constituent particles,
which after $n$ pairwise coalescence steps is $2^{n/3}$ times the
radius of the initial primary particles. The sintering becomes more slowly.

Using a continuously varying 
\begin{equation}
\tau(t)\propto (V/A(t))^4\ ,
\label{eq:Westerhoff}
\end{equation}
where $V$ is the constant solid volume of the aggregate and determines
$A_{\rm eq}$, the theory has recently been successfully applied 
to experimental sintering data of Ni-particles to calculate the
activation energy of the relevant diffusion
process, whereas the evaluation with a constant $\tau$ gives unreasonable
energy values.\shortcite{Tsyganov} 

\begin{figure}[t]
  \centering                           
  \epsfig{file=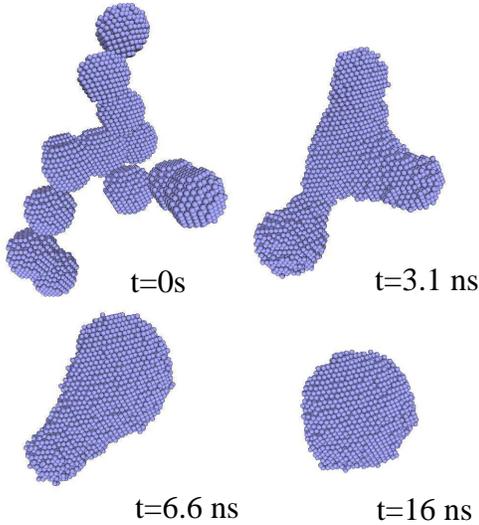,width=0.4\textwidth}      
  \caption{Snapshots during one simulation. The initial configuration consists of 12 clusters
  with 8372 atoms, $T=800$K. } 
  \label{fig:agglomerat}                        
\end{figure}                           

What is missing, is a microscopic justification of
Eq.~(\ref{eq:Koch-Friedlander}). Therefore we simulated the sintering
of agglomerates and measured the surface area.  Typical
snapshots are shown in Fig.~\ref{fig:agglomerat}.  We fitted the
surface area with the analytical solutions obtained for constant
$\tau$ or with Eq.~(\ref{eq:Westerhoff}), respectively. The constant 
$A_{\rm eq}$ was obtained as the average asymptotic surface area for
large times.

\begin{figure}
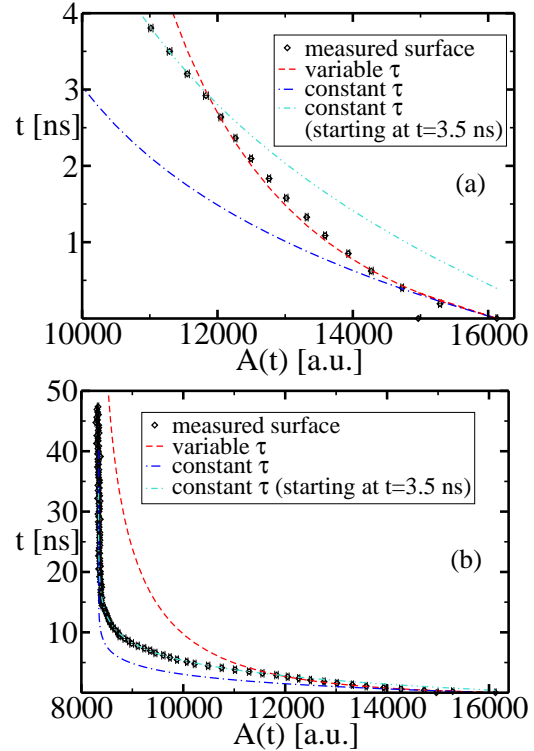

  \centering                           
  \epsfig{file=Figures/figure5.eps,width=0.43\textwidth}
  \epsfig{file=Figures/figure6.eps,width=0.43\textwidth} \\
  \caption{Fit with solution of (\ref{eq:Koch-Friedlander}) for constant
  and increasing   $\tau$ at the beginning (a) and the end (b) of the
  sintering.}           
  \label{fig:agg-fit}                        
\end{figure}

One can distinguish two different stages in which either a constant or
a variable $\tau$ gives a better description of the surface dynamics.
At early times when the particles successively merge together and the
average radius increases, the fit with a variable $\tau$ gives a
better description~(Fig.~\ref{fig:agg-fit}(a)). When the average diffusion
length does not increase significantly any more, a constant $\tau$ is a
better assumption~(Fig.~\ref{fig:agg-fit}(b)), but surprisingly its value is 
smaller than at the end of the early time regime with variable $\tau$. 

\subsection{Influence of elasticity on sintering}

In order to understand the influence of an external stress we allowed
elastic displacements of the atoms as described above.
We set up a configuration of two particles with periodic boundary
conditions in $x$-direction. The strain is controlled by the
total system size in $x$-direction, which is kept constant
during each simulation. 

\begin{figure}
  \centering                           
  \epsfig{file=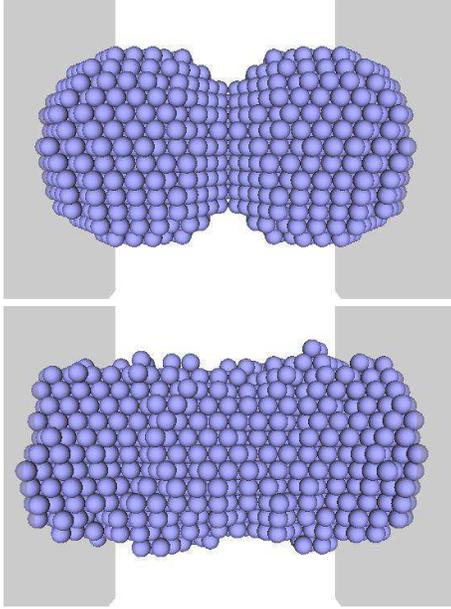,width=0.38\textwidth} 
  \caption{Sintering of two clusters with periodic boundary
  conditions in $x$-direction. An external stress (compressive/tensile)
  is applied by changing the system size in $x$-direction.}          
  \label{fig:pbcconf}                        
\end{figure}

In Fig.~\ref{fig:pbcconf} an initial and final configuration are
shown. One expects that the final configuration is reached faster
under compression than under tension.
We measure the average squared radius along the $x$ axis: 

\begin{equation}
R^2(t) = \frac{1}{N}\sum_{i=1}^{N}\left[(y_i - y_{\rm s})^2+(z_i -
  z_{\rm s})^2\right] 
\end{equation}
with
\begin{equation}
y_s = \frac{1}{N}\sum_{i=1}^{N}y_i\ \mbox{ and }\ 
z_s = \frac{1}{N}\sum_{i=1}^{N}z_i
\end{equation}

\begin{figure}[t]
  \centering                           
  \epsfig{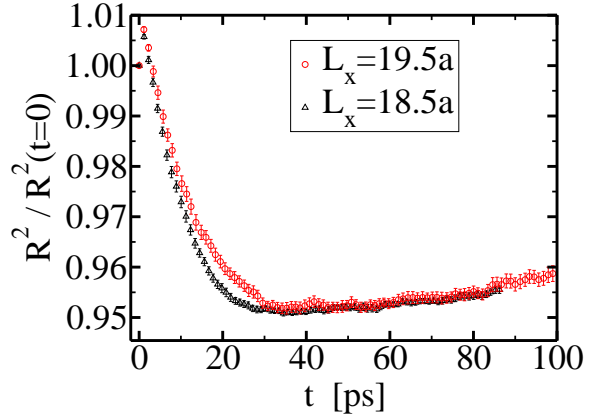} 
  \caption{Average squared radius perpendicular to  symmetry axis, scaled by
  initial value for the configuration shown in Fig.~\ref{fig:pbcconf}.} 
  \label{fig:pbc}                        
\end{figure}

The value $R^2(t)$ reflects how far away the system is from the
thermodynamic equilibrium state.
Its time dependence is plotted in
Fig.~\ref{fig:pbc}.  It is scaled with the initial value, $R^2(t=0)$, in
order to eliminate the effect of the Poisson ratio for a better
comparison of the curves for different initial strains. The initial
configuration (upper part of Fig.~\ref{fig:pbcconf}) are two spheres, whose
surface is not in thermal equilibrium. Therefore initially the curves
raise beyond 1 due to surface roughening.

Looking at the relaxation behaviour
we find that both systems evolve asymptotically similarly. However,
the relaxation process is faster for compressive strain which becomes
obvious from the larger slope in the beginning of the coalescence. The
equilibrium state is therefore reached earlier.
The reason is that $\alpha>1$ in Eq.~(\ref{eq:Ealj_stable}) for
hopping diffusion in Lennard-Jones-Systems, and the binding of atoms
on the surface becomes weaker under lateral compression of the surface
\shortcite{Schroeder}. This implies that the activation energy
decreases under lateral compression.

The cylindrical
configuration reached at the end of our simulation probably is only
metastable: Due to the periodic boundary conditions it corresponds to
an infinitely long solid cylinder which should undergo a Rayleigh
instability and split up into separate spheres. We think that thermal
fluctuations in this direction are the reason why $R$ increases again
for larger times.

\section{CONCLUSIONS AND OUTLOOK}

We developed a 3d-KMC simulation program to model the sintering of
nano-particles. The special feature of the program is that atoms are not
fixed at their lattice sites in contrast to common KMC codes. This
allows us to analyse the effect of elasticty on the sintering behavior.

The validation of the Koch-Friedlander theory leads to interesting
results: The assumption of a constant $\tau$ can only be applied if
the characteristic diffusion length does not change.
At the beginning of the sintering process of an agglomerate consisting
of {\em many} clusters the  characteristic length scale changes. In
this case the assumption of a variable $\tau$ gives a better
description of the surface evolution.

We found that the equilibration time for two clusters of different size
follows the $r^4$ power law with the reduced radius (\ref{eq:Sinterzeit}). 

The influence of elasticity for our set of parameters is only felt in
the case of an external force. There it is found that compressive stress
leads to faster relaxation. Without external stress it seems that the
influence of the surface tension is  negligible. This needs not
always be the case, as our potential does not show strong surface tension. 

Therefore a next step would be to choose a potential
that shows a stronger surface tension. So far grain
boundaries were neglected as we used 
a continuous fcc lattice. In the current development of the code these
grain boundaries will be implemented.\\

\section*{ACKNOWLEDGMENT}
This work is supported by the Deutsche Forschungsgemeinschaft within
SFB 445, ``Nano-particles from the gas phase''. 

\bibliographystyle{chikako}      
\bibliography{mybib} 

\end{document}